\documentstyle[12pt,a4,amssymb,epsfig]{article}
\begin{document}

\begin{flushright}
MPI-PhT/97-37, LMU-07/97\\
September 1997\\
\end{flushright}
\begin{center}
\large {\bf A low--energy compatible SU(4)--type Model for 
Vector Leptoquarks of Mass $\leq$ 1 TeV}
\\
\mbox{ }\\
\normalsize
\vskip1cm
{\bf Andreas Blumhofer}
\vskip0.2cm
Department of Physics, University of Munich\\
Theresienstrasse 37, D--80333 Munich \\
\vskip0.5cm
{\bf Bodo Lampe}
\vskip0.2cm
Max Planck Institute for Physics \\
F\"ohringer Ring 6, D--80805 Munich \\
\vspace{1cm}

{\bf Abstract}\\                          
\end{center}                              

The Standard Model is extended by a $SU(2)_L$ singlet 
of vector leptoquarks. An additional $SU(4)$ gauge symmetry between 
right--handed up quarks and right--handed leptons is introduced 
to render the model 
renormalizable. The arrangement is made in such a way that 
no conflict with low energy restrictions is encountered. 
The $SU(2)_L$ singlet mediates interactions between the 
right--handed leptons and up type quarks for which only moderate 
low energy restrictions $M_{LQ}/g_{LQ} > $ few hundred GeV exist. 
However, it is {\it not} a candidate 
to explain the anomalous HERA data at large $Q^2$ 
because theoretical reasons imply that $g_{LQ} \geq g_s$ which 
would give a much stronger anomalous HERA effect. 
We furthermore argue that the inequality $g_{LQ} \geq g_s$ is a general 
feature of consistent vector leptoquark models.
Although our model is not relevant for HERA, it is 
interesting per se as a description of leptoquarks of mass $\leq$ 1 TeV 
consistent with all low--energy requirements.  

\vspace{1cm}

{\bf Introduction. }
There has been an increasing interest in leptoquarks of mass 
$M_{LQ}\sim$ few hundred GeV in 
the last months, due to the exciting possibility of observing 
such particles at HERA \cite{hera,altarelli,kalinowski,barger}. 
Single leptoquarks may be produced 
in electron proton collisions directly in an s--channel process 
$eq \rightarrow LQ$
whereas in proton--proton collisions they contribute more indirectly 
via t--channel exchange or are pair--produced. 
Correspondingly, 
leptoquarks seen at HERA 
need not necessarily satisfy the Tevatron bounds \cite{teva}, 
because the processes and couplings involved are different.

On the theoretical side there are many extensions of the standard 
model which predict the existence of leptoquarks with masses 
which could be of the order of a few hundred GeV 
\cite{pati,fritzsch}. 
However, phenomenological considerations \cite{sacha,shanker,leurer} 
show that most of these 
models are in conflict with low energy data. Either they induce 
proton decay or various FCNC processes or they enhance leptonic 
decays of pseudoscalar mesons. The phenomenological restrictions 
can usually be expressed in terms of the ratio $M_{LQ}/g_{LQ}$ 
where $g_{LQ}$ is the coupling of the leptoquarks to quarks and 
leptons.   

{\bf Low Energy Constraints. }
More in detail, the processes that lead to the strongest 
bounds on the leptoquark mass and couplings are  
\\
(i) {\it proton decay} \\ 
This is induced when the leptoquark 
has diquark couplings as well, so that processes 
$qq\rightarrow ql$ etc are possible. 
Leptoquarks of this type are out of reach of any future 
collider. 
\\
(ii) {\it flavor changing neutral current processes} \\ 
These are induced when the leptoquark couples to more than one  
generation in the lepton and/or the quark sector. The 
strongest bound arises from the decay 
$K_L \rightarrow \mu e$ which is induced by exchange 
of a leptoquark in the t--channel and typically given by 
$M_{LQ}/g_{LQ} \geq 100$ TeV \cite{sacha}. 
\\
(iii) {\it leptonic decays of pions and other pseudoscalars} \\
This bound is particularly strong for leptoquarks that 
couple both to left--handed and right--handed quarks, 
namely $M_{LQ}/g_{LQ} \geq 100$ TeV \cite{sacha}.
\\
(iv) {\it other processes} \\
There are a number of other processes like D--decays, 
$K^0\bar K^0$ mixing, $\mu$--decays, $\tau$--decays 
\cite{sacha} and    
atomic parity violation \cite{leurer,babu}, all of which give 
weaker constraints to leptoquark masses and couplings 
and are compatible with a leptoquark $M_{LQ}/g_{LQ} \sim O(1)$ TeV.
Thus, these processes are not in contradiction with a 
low lying leptoquark of mass $\sim 200$ GeV provided the 
leptoquark coupling is sufficiently weak and not of the 
order of the strong coupling constant.  
The most interesting among these restrictions are perhaps the 
atomic experiments, because leptoquarks give parity violating 
contributions like ${g_{LQ}^2\over M_{LQ}^2} 
(\bar e \gamma_{\mu}\gamma_5 e)
(\bar q \gamma_{\mu} q) $
to the interactions of electrons and quarks in ordinary atoms \cite{leurer}. 

In the model presented below the bounds (i-iii) are avoided. 
There are no proton decays and the FCNC processes involving Kaons 
are avoided by 
the leptoquarks coupling to up-- but not to down--type quarks. 
The restrictions from 
D--decays are much less severe than from 
K--decays, typically given by
$M_{LQ}/g_{LQ} \geq O(1)$ TeV \cite{sacha}. 
Furthermore, the strong bound from pion decays is avoided 
because the leptoquarks 
couple chirally, and in particular they couple 
only to right--handed quarks.
Note also that there is no "CKM--type" mixing in the model. 

{\bf General Analysis. }
Possible leptoquarks interactions have been analyzed in ref. 
\cite{buchm1} in a model independent way from a purely 
phenomenological point of view. Because of their coupling 
to quarks, all leptoquarks carry color (in the fundamental 
representation). Furthermore, all leptoquark fields have 
dimension 1 and integer spin (0 or 1), i.e.  
they are either scalar or vector fields. 
Depending on whether they interact with a fermion--antifermion 
or a fermion--fermion system they carry fermion number 
$F=3B+L=0$ or $-2$. 
Using the assumption 
that the leptoquark interactions respect the symmetries of the 
standard model the most general effective Lagrangian involving 
leptoquarks was derived in \cite{buchm1}. 
These relatively mild assumptions lead to a variety 
of leptoquarks. In the model presented below just one of these is 
selected by requiring the following principles  
\begin{itemize}    
\item {\it gauge principle} \\ 
this assumes that the leptoquarks themselves arise as vector bosons 
of a new gauge group. This excludes all scalar leptoquarks from the list.  
Indeed, leptoquarks are naturally gauge bosons and 
as such they appear in many extensions of the standard model. 
Scalar leptoquarks appear in conjunction with the corresponding 
Higgs mechanism (see below) 
or as superpartners in supersymmetric theories \cite{rparity,
altarelli,kalinowski,barger}. 
It is true that Tevatron data seem to exclude vector 
leptoquarks below 300 GeV \cite{teva}. 
Therefore, 
in this article 
we explore the theoretical possibility of a vector leptoquark 
of mass 300 GeV $\leq m_{LQ} \leq$ 1 TeV.   
\item {\it universality} \\ 
this implies that the leptoquark couplings to all families are the same. 
It is a reasonable assumption in view of the known universality 
of all the other gauge interactions.  
\item {\it vanishing fermion number} \\ 
this leaves only the $F=0$ 
leptoquarks in the list. In the appendix there will be 
a short discussion about what happens if this assumption is given 
up. It turns out, that one can construct a consistent 
$F=-2$ vector leptoquark model 
based on the gauge group 
$SU(5) \times SU(3)  \times SU(2) \times U(1)$, cf. the appendix. 
\end{itemize}    

Together with the low energy constraints, 
these principles are so strong that only one of the leptoquarks 
passes the requirements, namely an $F=0$ vector particle $V_{\mu}^i$ 
with quantum numbers 
$(3,1,{5\over 3})$ under $SU(3)_c \times SU(2)_L \times U(1)_Y $ 
and lepton--quark interactions 
$g_{LQ} \bar u_R^i \gamma^{\mu} e_R  V_{\mu}^i$ +c.c. (i=color index). 
A goal of this letter is to embed this particle and its 
interaction into a renormalizable (and thus consistent) extension  
of the standard model. 
A Higgs 
mechanism will be invoked to obtain the leptoquark mass.    
The minimal extension of the standard model which includes the 
fields $V_{\mu}^i$ is by an $SU(4)$ gauge group which acts on the 
right--handed 
quartet $p_R$ formed by $e_R$ and $u_R^i$, i=1,2,3. $V_{\mu}^i$ are thus 
leptoquarks of the Pati--Salam type, but without interactions to d--type 
quarks and to left--handed fermions. 
The total symmetry group of the model is
$SU(4) \times SU(3)  \times SU(2)_L \times U(1)_X $. 
Let us write down the Lagrangian: 
\begin{eqnarray} \nonumber
L&=&\bar p_R i\gamma^{\mu} [ \partial_{\mu}+ig_1 X(p_R)C_{\mu} 
                           +ig_4 R_{\mu}^a {\rho_a \over 2} ] p_R 
\\ \nonumber & & 
+\bar d_R i\gamma^{\mu} [ \partial_{\mu}+ig_1 Q(d_R)C_{\mu}
   +ig_3 L_{\mu}^a {\lambda_a \over 2}] d_R
\\ \nonumber & & 
+\bar l_L i\gamma^{\mu} [ \partial_{\mu}+ig_1 Y(l_L)C_{\mu} 
                           +ig_2 W_{\mu}^a {\tau_a \over 2} ] l_L 
\\ \nonumber & & 
+\bar q_L i\gamma^{\mu} [ \partial_{\mu}+ig_1 Y(q_L)C_{\mu} 
                           +ig_2 W_{\mu}^a {\tau_a \over 2} 
       +ig_3 L_{\mu}^a {\lambda_a \over 2}] q_L
\\ 
& & -{1\over 4} R_{\mu \nu}^a R^{\mu \nu a}
-{1\over 4} L_{\mu \nu}^a L^{\mu \nu a} 
-{1\over 4} W_{\mu \nu}^a W^{\mu \nu a}  
-{1\over 4} C_{\mu \nu} C^{\mu \nu }  
\label{4119}
\end{eqnarray}
Here $q_L$ and $l_L$ are the left--handed quark and lepton
doublets, and $e_R$, $d_R$ and $u_R$ the right--handed charged
leptons, down-- and up--type quarks, respectively. Color,
weak isospin and generation indices have been suppressed.
$X(p_R)$ is the U(1) charge of the quartet 
$p_R=(u_R,e_R)$. It will 
be fixed later to be $X(p_R)={1\over 4}$ 
by requiring that the electromagnetic coupling comes 
out right. For all other fermion fields, $l_L$, $d_R$ and $q_L$, 
the X--charge agrees with the weak hypercharge. $C_{\mu}$ is the 
$U(1)_X$ gauge field which will mix with the other neutral fields 
of the model, $W_{\mu}^3$ and $R_{\mu}^{15}$. $R_{\mu}$, $L_{\mu}$ 
and $W_{\mu}$ are the gauge bosons of the $SU(4)$, $SU(3)$ and 
$SU(2)$, respectively, with gauge couplings $g_4$, $g_3$ and $g_2$.  
The algebra of $SU(4)$ is spanned 
by the matrices $\rho_a$, a=1,...,15, where $\rho_1$,...,$\rho_8$ 
are the SU(3) $\lambda$--matrices.   
For example, $\rho_{15}$ is given by 
${\rho_{15} \over 2}= {1\over \sqrt{24}}$diag$(-1,-1,-1,3)$. 
One can write 
\begin{equation}
R_{\mu}=R_{\mu}^a {\rho_a \over 2}
= {1\over \sqrt{2}} 
\pmatrix{\hat R_{\mu} -{S_{\mu}\times 1\over \sqrt{12}}  &V_{\mu}\cr
        V^+_{\mu} &\sqrt{3\over 4}S_{\mu} \cr} \, ,
\label{557}        
\end{equation}     
where group indices have been suppressed. One sees that besides the 
leptoquarks $V_{\mu}$, there is an octet $\hat R_{\mu}$ and a singlet 
$S_{\mu}:=R_{\mu}^{15}$ of vector bosons. 
$\hat R_{\mu}$ will mix with the SU(3) 
octet $L_{\mu}$ to form 8 massless gluons and 8 massive 
'axigluon' states. 

The quantum number assignments 
for the fermions can be found in Table 1.   
A family symmetry is assumed which is only broken by 
fermion mass terms. 
As can be seen from Table 1, there are big differences 
as compared to the Pati--Salam model. The main difference is 
that not all possible leptons and quarks are put 
into a $SU(4)$ multiplet, but only the right--handed electron 
and up--quark.  
Note that this is a {\it chiral} model, i.e. left-- and right--handed 
fermions  
behave non--symmetric, so as to embed the weak interactions in 
the model.  

\begin{table} 
\label{tab1}  
\begin{center}
\begin{tabular}{|c|c|c|c|c|}
\hline
 & $SU(4)$ & $SU(3)$ & $SU(2)_L$ & $U(1)_X$ \\
\hline
$q_L$ & 1 & 3 & 2 & ${1\over 6}$ \\
$l_L$ & 1 & 1 & 2 & $-{1\over 2}$ \\
$p_R$ & 4 & 1 & 1 & ${1\over 4}$ \\
$d_R$ & 1 & 3 & 1 & $-{1\over 3}$ \\
\hline
\end{tabular}
\bigskip
\caption{Quantum number assignments}
\end{center}
\end{table}

Higgs--terms have been omitted in the Lagrangian. 
They will be discussed later, and as usual, they will provide boson 
and fermion masses. Essentially, there will be one Higgsfield, 
$H(\bar 4,3,1)$ with vev $v$, 
which will break $SU(3) \times SU(4)$ to color $SU(3)$, 
and three other Higgs fields, which break $SU(2)_L$ and give 
masses to the fermions. 
The leptoquark masses are then 
of the order $g_4 v$ whereas the W and Z mass turn out to be 
as in the Standard Model.  

{\bf Color Sector. } 
Those Higgs interactions should break the $SU(3) \times SU(4)$ 
symmetry down to the diagonal $SU(3)_c$ in a similar fashion 
than happens in the so--called chiral--color models \cite{frampton} 
which are based on $SU(3)_L \times SU(3)_R$. 
In those models, the gauge fields 
$\hat R_{\mu}$ and $L_{\mu}$ couple to right-- and left--handed  
currents and are rotated in order to get the QCD couplings 
to the gluons $G_{\mu}$ right, 
\begin{eqnarray} \nonumber
L_{\mu}=c_{\theta} N_{\mu}+s_{\theta} G_{\mu} 
\\
\hat R_{\mu}=-s_{\theta} N_{\mu}+c_{\theta} G_{\mu}
\label{51}
\end{eqnarray}  
The 'right--handed' bosons are written with a hat here in order to 
make the analogy with our model clear. 
$N_{\mu}$ are the 'axigluons' which have to become heavy by
a suitable Higgs mechanism which breaks $SU(3)_L \times SU(3)_R$ 
to the diagonal $SU(3)$. In the limit that the $SU(3)_L$ and the 
$SU(3)_R$ gauge couplings are identical, the $N_{\mu}$ couple 
purely axially to fermions. Hence the name axigluons. 
Present Tevatron restrictions on 
axigluons are such that an axigluon with mass 500 GeV would 
be compatible with almost all bounds \cite{abeaxi,bagger}. 

In the SU(4) case at hand one can proceed analogously. 
The relevant interactions of the quarks are given by 
\begin{equation}
L_{strong}=g_3 \bar d_R \gamma_{\mu}L^{\mu}d_R +
g_3 \bar d_L \gamma_{\mu}L^{\mu}d_L +
g_3 \bar u_L \gamma_{\mu}L^{\mu}u_L +
g_4 \bar u_R \gamma_{\mu}\hat R^{\mu}u_R    \,  . 
\label{5572}
\end{equation}
Inserting Eq. (\ref{51}), one can prove that the ordinary gluon 
interactions are reproduced if 
\begin{equation}
g_s=g_3 s_{\theta}=g_4 c_{\theta}
\label{5573}
\end{equation}
or, equivalently, 
\begin{equation}
g_s^{-2}=g_3^{-2}+g_4^{-2}  \,  .
\label{5574}
\end{equation}
An immediate consequence of these relations is that 
both $g_3$ and $g_4$ are necessarily larger than  
the QCD coupling, i.e. for scales below 1 TeV one 
has $g_{3,4} \gtrsim O(1)$. This is undesired in view 
of the anomalous DESY--HERA data because the leptoquark 
coupling to fermions is $g_{LQ}=g_4$ and the HERA data 
require a smaller coupling. Furthermore, one has 
low--energy constraints $M_{LQ}/g_{LQ} \gtrsim O(1)$ TeV which 
imply that the leptoquark mass $M_{V}$ in our model is closer to 1 TeV 
than to 200 GeV.  
It should be stressed that the relation $g_{LQ} \gtrsim O(1)$  
is a characteristic feature of the class of models 
discussed in this letter. 

{\bf Higgs Sector. }
In principle, there is some ambiguity in choosing the Higgs 
multiplets and the Higgs potential. In the following we present 
one possibility which is consistent with the 
Standard Model at low energies. 
The Higgs fields in this scenario have a rather 
complicated structure. From the quantum number assignments in Table 
1 it is evident that one needs different Higgs multiplets for the various 
fermions to become heavy. There is a Higgs field   
$H_e(\bar 4,1,2,-{1\over 2}-X(p_R))$ with vev 
$v_e\delta_{\bar \alpha 4}\delta_{i2}$ for the lepton 
masses, a Higgs field 
$H_u(\bar 4,3,2,{1\over 6}-X(p_R))$ with vev 
$v_u\delta_{\bar \alpha \beta}\delta_{i1}$ 
for the up quark  
masses and a Higgs field 
$H_d(1,1,2,{1\over 2})$ with vev $v_d\delta_{i2}$ for the down quark   
masses. Here, $\bar \alpha$, $\beta$ and $i$ are $SU(4)$, 
$SU(3)$ and $SU(2)_L$ indices, 
respectively. Altogether, the Yukawa terms of the Lagrangian 
are given by 
\begin{equation}
L_{Yuk}=h_e H_e \bar l_L p_R +h_u H_u \bar q_L p_R
+h_d H_d \bar q_L d_R+c.c.
\label{5571}
\end{equation}
It is interesting to note that apart from giving masses 
to the fermions, part of these interactions have the form 
of interactions of scalar leptoquarks. However, due to the smallness 
of the Yukawa couplings, they are relevant (if at all) only for the third 
family. Of course, Eq. (\ref{5571}) induces other interactions 
between fermions and Higgs fields as well. 
If the Higgs masses 
are not too large, there may be some relevance for interactions 
of the top quark, because the top quark Yukawa coupling is not 
small. For example, in Eq. (\ref{5571}) 
there are colored Higgs fields which mediate 
interactions between bottom-- and top--quark.  

Note further, that $H_d$ has the quantum numbers of the Standard Model 
Higgs field. 
A reasonable outcome for the vacuum expectation values of 
the three Higgs fields would certainly be 
$v_u : v_d : v_e \sim m_t : m_b : m_{\tau}$. However, it will be 
shown that this is not compatible with the requirement that 
$\gamma$ and Z couplings to fermions are as given in the 
Standard Model. In fact, to achieve this goal, we will be forced 
to introduced yet another Higgs multiplet, $H(\bar 4,3,1,{5\over 12})$ 
with vev $v\delta_{\bar \alpha \beta}$ which 
breaks $SU(4) \times SU(3)$ to $SU(3)_c$ 
\footnote{The X--charge ${5\over 12}$ of H  is fixed by 
the requirement that the components $H_{\bar \beta \beta}$, 
$\beta =1,2,3$ are neutral.}. 
Although $H_u$ 
in principle does the same job, it is incompatible with the 
correct Z couplings to fermions (see below). 
A typical solution will be that $v$ is of the order of the 
leptoquark mass, and $v_d$, $v_u$ and $v_e$ of the order of the electroweak 
symmetry breaking scale or somewhat smaller. $v_d$ tends 
to be the largest among $v_d$, $v_u$ and $v_e$, thus 
playing approximately the role of the Standard Model vev. 

The Higgs vevs induce the following vector boson masses : 
\begin{itemize}
\item Axigluon mass : ${1\over 2}m_N^2={1\over 2}
(g_3^2+g_4^2)(v^2+v_u^2)$ \\ 
This mass is always large because it involves the large 
vev $v$ and both strong couplings $g_3$ and $g_4$. 
\item Leptoquark mass : ${1\over 2}m_V^2={1\over 4}   
g_4^2(v^2+v_u^2+v_e^2)$  \\
This mass is in general smaller than the axigluon mass 
because a term $\sim g_3^2$ is missing (assuming $v_e << v$). 
However, it competes with the mass of the neutral boson 
related to the $\rho^{15}$ generator, as discussed below. 
\item Mass matrix of the neutral vector bosons $C_{\mu}$, 
$W^3_{\mu}$ and $S_{\mu}$ :  ${1\over 2} M^2=$ \\ 
\begin{equation}
\pmatrix{
{g_1^2\over 4}({25\over 12} v^2+{3\over 36} v_u^2+ v_d^2+
{9\over 4} v_e^2) & 
{g_1g_2\over 4}(-{1 \over 2} v_u^2- v_d^2+
{3\over 2} v_e^2) &
{g_1g_4\over 8\sqrt{6}}( 5v^2- v_u^2+9 v_e^2) \cr
{g_1g_2\over 4}(-{1 \over 2} v_u^2- v_d^2+
{3\over 2} v_e^2) &
{g_2^2\over 4}(3 v_u^2+ v_d^2+ v_e^2) & 
{3g_2g_4\over 4\sqrt{6}}( v_u^2+ v_e^2) \cr
{g_1g_4\over 8\sqrt{6}}( 5v^2- v_u^2+9 v_e^2) &
{3g_2g_4\over 4\sqrt{6}}( v_u^2+ v_e^2) & 
{g_4^2\over 8}( v^2+ v_u^2+3 v_e^2) 
 \cr} 
\label{5577}        
\end{equation} 
Note that this matrix has one vanishing and two nonvanishing 
eigenvalues. The corresponding eigenstates will be called 
$A_{\mu}$, $Z_{\mu}$ and $T_{\mu}$. By calculating the 
characteristic polynomial, one sees that the mass of the state $T_{\mu}$ 
is governed by the vev $v$ 
whereas $m_Z$ is independent of $v$ and thus smaller. 
The rotation matrix which diagonalizes ${1\over 2} M^2$ will be 
called r, i.e. 
$r^T {1\over 2} M^2 r=$diag$(0,{1\over 2}m_Z^2,{1\over 2}m_T^2)$. 
\item $W^{\pm}$ mass : ${1\over 2}m_W^2={1\over 4}g_2^2
(3v_u^2+v_e^2+v_d^2)$ \\
This is like in the Standard Model with $v_{SM} \approx 175$ GeV 
replaced by $3v_u^2+v_e^2+v_d^2$. When diagonalizing the mass 
matrix of the neutrals, it will be possible to 
maintain the relation $m_W=c_Wm_Z$ where $c_W$ is the 
cosine of the ordinary Weinberg angle.
\end{itemize}

{\bf Electroweak Sector. }
The mass matrix of the neutral gauge bosons has to fulfill two 
requirements. First of all, the Z--mass must come out as 
${1\over 2}m_Z^2={e^2\over 4s_W^2c_W^2} (3v_u^2+v_e^2+v_d^2)$ to 
fulfill $m_W=c_Wm_Z$. Secondly, the couplings of photon and Z ,which 
are linear combinations of $C_{\mu}$,
$W^3_{\mu}$ and $S_{\mu}$, must be as in the standard model. For 
example, the coupling of the Z to left--handed fermions $f_L$ must 
be ${e\over s_Wc_W}(T^3_f-s_W^2 Q_f)$. In these relations, the 
three quantities $e$, $s_W=\sqrt{1-c_W^2}$ and $Q$ still have to 
be defined in the framework of our model. $s_W$ is simply 
$s_W={e\over g_2}$, as in the standard model. $e$ and $Q$ are 
properties of the massless photon state $A_{\mu}$ and given by 
\begin{equation}
{e^2\over g_1^2}+{e^2\over g_2^2}+{25\over 6}{e^2\over g_4^2}=1
\label{507}
\end{equation}
and 
\begin{equation}
Q=X-{5\over \sqrt{6}} T_{15} +T_3
\label{50}
\end{equation}
because the photon state is given by 
\begin{equation}
{A_{\mu} \over e} = {C_{\mu} \over g_1}+{W^3_{\mu} \over g_2} 
-{5 \over \sqrt{6}} {S_{\mu} \over g_4}
\label{49}
\end{equation}
Note that $T_{15}={\rho_{15} \over 2}$ and $T_3={\tau_{3} \over 2}$. 
Furthermore, the coupling $g_Y$ corresponding to the Standard 
Model weak hypercharge is given by 
\begin{equation}
g_Y^{-2}=g_1^{-2}+{25\over 6} g_4^{-2}  \, . 
\label{491711}
\end{equation}

Given $e=0.303$ and $g_2=0.636$, Eq. (\ref{507}) can be 
considered as a relation between $g_1$ and $g_4$. One may 
introduce an angle $\phi$ by the relations 
\begin{equation}
c_{\phi}={e\over g_1 c_W} \qquad s_{\phi}={5e\over \sqrt{6}g_4c_W}
\label{491}
\end{equation}
with $c_{\phi}^2+s_{\phi}^2=1$. 
This angle will simplify the notation in the following. 

The requirement that the couplings of photon and Z 
must be as in the standard model, completely fixes the 
rotation matrix r which must be used to diagonalize 
${1\over 2} M^2$, Eq. (\ref{5577}). It can be shown that 
all photon and Z couplings come out in agreement with the 
Standard Model if and only if one has 
\begin{equation}
r=\pmatrix{c_Wc_{\phi} & -s_Wc_{\phi} & s_{\phi} \cr
             s_W        & c_W  & 0  \cr
          -c_Ws_{\phi} & s_W s_{\phi} & c_{\phi} \cr}
\label{493}
\end{equation}
In general this matrix $r$ will {\it not} diagonalize 
${1\over 2} M^2$, and thus in general the Standard Model 
couplings cannot be reproduced. However,  there is one simple 
condition under which $r$, Eq. (\ref{493}), completely 
diagonalizes ${1\over 2} M^2$ and at the same time gives the correct Z 
mass, namely 
\begin{equation}
v_u^2+v_e^2={2\over 5} s_{\phi}^2(3v_u^2+v_e^2+v_d^2)
={4\over 5e^2} s_{\phi}^2s_W^2c_W^2m_Z^2
\label{494}
\end{equation}
This condition can be fulfilled for various values of the vevs. 
A typical solution is $v$ being larger than $v_d$ and this in turn 
is larger than $v_{u}$ and $v_{l}$. However, it 
should be noted that one must not take $v_u=v_e=0$, because 
otherwise $g_4 \rightarrow \infty$ according to Eq. (\ref{494}).
Within this solution, the leptoquark 
mass is always of the same order as the mass of the T--particle 
(to within $\pm 50$ GeV), whereas the axigluon masses can be made 
higher than 1 TeV (if desired). Note that there are Tevatron 
limits on the mass of neutral vector bosons, of about 550 GeV 
\cite{ima}. This fact can be used to argue that 
the leptoquark in our model should have mass $m_V \gtrsim 500$ GeV.    

Due to the mixing of photon, Z and T, 
there is another constraint 
on the coupling $g_4$ (weaker than $g_4 \geq g_s$), 
arising from the limit $s_{\phi} \leq 1$ 
in the electroweak sector. According to Eq. it is 
given by $g_4 \geq {5e\over \sqrt{6}c_W } \approx 0.70$. 
Conversely, the condition $g_4 \geq g_s$ translates into 
a constraint for $s_{\phi}$, namely $s_{\phi}^2 \lesssim 0.5$. 

It might be interesting to examine the Higgs content of the 
various vector bosons, for simplicity in the limit 
$v >> v_d >> v_{u,l}$. In that limit 3 of the 4 real 
components of $H_d$ are eaten up by W and Z leaving 
the Standard Model Higgs field as a real particle. 
The longitudinal components of the axigluons, leptoquarks 
and neutral (8+6+1=15 real degrees of freedom) are 
given by 15 components of the Higgs multiplet 
$H(\bar 4,3,1)$, namely Im $H_{\bar \alpha \beta}$ 
($\alpha ,\beta =1,2,3$), $H_{\bar 1 \beta}$ 
($\beta =1,2,3$) and Im $H_{\bar \beta \beta}$, 
respectively. The 9 remaining real parts 
of $H_{\bar \alpha \beta}$ and $H_{\bar \beta \beta}$ 
will be real Higgs particles.

{\bf Checking the Vector Boson Self Interactions. }
Besides its interactions with leptons and quarks, the leptoquark 
interacts with other Standard Model particles, namely with 
the photon, the gluon and the Z. This can be seen by 
working out the terms 
$-{1\over 4} R_{\mu \nu}^a R^{\mu \nu a}
-{1\over 4} L_{\mu \nu}^a L^{\mu \nu a}$
in the Lagrangian. Not surprisingly, one finds  
that the coupling strength to the photon is ${5\over 3}e$ 
and to the gluons is given by $g_s$. 
The coupling to gluons is being used in the Tevatron 
searches for leptoquarks via the process 
$q\bar q \rightarrow g^{\ast} \rightarrow V \bar V$. 
Note that the Yang--Mills terms induce several other 
interesting vector boson self couplings which will 
not be discussed here. 

\begin{table} 
\label{tab2}  
\begin{center}
\begin{tabular}{|c|c|c|c|c|c|c|}
\hline
 & $X^3$ & $X\tau^2$ & $\rho^3$ & $\lambda^3$ & $X\rho^2$ & $X\lambda^2$ \\
\hline   
$q_L$ & ${1\over 36}$ & ${1\over 4}$ & 0 & 2 & 0 & ${1\over 6}$ \\
$l_L$ & $-{1\over 4}$ & $-{1\over 4}$ & 0 & 0 & 0 & 0 \\
$p_R$ & $-{1\over 16}$ & 0 & $-1$ & 0 & $-{1\over 8}$ & 0 \\
$d_R$ & ${1\over 9}$ & 0 & 0 & $-1$ & 0 & ${1\over 6}$  \\    
\hline 
& $-{25\over 144}$ & 0 & $-1$ & 1 & $-{1\over 8}$ & ${1\over 3}$ \\ 
\hline
\end{tabular}   
\bigskip
\caption{Anomalies of the model. The generators of $SU(4)$, $SU(3)$, 
$SU(2)_L$ and $U(1)_X$ are denoted by $\rho$, $\lambda$, $\tau$ and $X$, 
respectively. Note that the anomaly for $\tau^3$ vanishes as a 
consequence of a general $SU(2)$ property, 
and the anomalies for $X^2\tau$, $X^2\lambda$ and 
$X^2\rho$ are zero due to the tracelessness of the $SU(2)$, $SU(3)$ 
and $SU(4)$ generators.  }
\end{center}
\end{table}

{\bf Anomaly Cancellation and Unification. } It is well known that 
in the Standard Model all $\gamma_5$--anomalies cancel. In the present 
model this does not happen, unless additional exotic fermion multiplets 
are introduced. As is shown below, this works similarly as  
in the chiral--color models based on $SU(3)_L \times SU(3)_R$ 
\cite{rajpoot}. 

The list of anomaly coefficients of the standard fermions 
is given in Table 2. They are summed up in the last line 
of this table.    
Additional fermion multiplets have to be chosen in such a way 
that their contributions exactly cancel the numbers in the last 
line of Table 2. 
We have scanned through all possible fermion representations 
of $SU(4)\times SU(3)  \times SU(2)_L \times U(1)_X $ and have 
obtained a very simple solution to this problem in the form 
of three additional fermion multiplets, namely a 
left--handed $SU(4)$ quartet 
$F_L(4,1,1,{1\over 4})$, a $SU(3)$ triplet $G_R(1,3,1,{2\over 3})$ 
and a singlet state $K_R(1,1,1,-1)$. Note that $F_L$ and $G_R$ 
have similar but not identical quantum numbers to the standard 
fermions $p_R$ and $d_R$, respectively. In Table 3 the contributions  
of these new fermions to the various anomaly coefficients are given. 
The last line of the table sums up these contributions. By comparing 
the last lines of Tables 2 and 3 it can be seen that all the 
anomalies completely cancel. 
It should be noted that the additional fermions have been chosen 
to be singlets under $SU(2)_L$. This makes sure that the cancellation 
of the $SU(2)_L$ anomalies (second row of Table 2) is not 
disturbed. Note further, that a family repetition structure of the 
standard fermions and of the new fermions is understood in all 
the considerations. 

\begin{table} 
\label{tab3}  
\begin{center}
\begin{tabular}{|c|c|c|c|c|c|c|}
\hline
 & $X^3$ & $X\tau^2$ & $\rho^3$ & $\lambda^3$ & $X\rho^2$ & $X\lambda^2$ \\
\hline   
$F_L$ & ${1\over 16}$ & 0 & 1 & 0 & ${1\over 8}$ & 0 \\
$G_R$ & $-{8\over 9}$ & 0 & 0 & $-1$ & 0  & $-{1\over 3}$ \\
$K_R$ & 1             & 0 & 0 & 0 & 0 & 0  \\    
\hline 
& ${25\over 144}$ & 0 & 1 & $-1$ & ${1\over 8}$ & $-{1\over 3}$ \\ 
\hline
\end{tabular}   
\bigskip
\caption{Anomaly contributions of the new fermions. The same notation 
as in Table 2 is used. 
  }
\end{center}
\end{table}

The next step is to generate mass terms for the new fermions. 
These masses have to be large enough to avoid conflict with 
existing bounds on heavy fermions. Since $F_L$, $G_R$ and $K_R$ 
are $SU(2)_L$ singlets, the particularly strong constraints on 
$SU(2)_L$ doublets do not apply \cite{pdb}. 
Therefore, masses of the new 
fermions larger than 500 GeV are certainly compatible with all present 
limits \cite{pdb}. In the following we want to describe how 
masses in the range between 500 and 1000 GeV can be obtained. 
The singlet property of $F_L$ and $G_R$ under $SU(2)_L$ has 
the convenient consequence, that it allows to write down a 
Yukawa coupling term 
of the form $\bar G_R F_L H(\bar 4,3,1,{5\over 12})$,  
where $H(\bar 4,3,1,{5\over 12})$ is the Higgs multiplet with vev 
$v\delta_{\bar \alpha \beta}$ 
used earlier to break $SU(4) \times SU(3)$ to $SU(3)_c$. This 
Yukawa term will give a mass to $G_R$ and the first three components  
of $F_L$ which is of the order 
$v$, i.e. the scale of the leptoquark mass. 
To obtain a mass term for $K_R$ and the fourth 
component of $F_L$, an additional 
Higgs field $H_K(\bar 4,1,1,-{5\over 4})$ with vev 
$v_K\delta_{\bar \alpha 4}$ has to be introduced. 
$v_K$ must be of the same order of magnitude as $v$ if all 
the new fermions are to be (at least) as heavy as the leptoquarks. 
Thus, the  
Yukawa interactions of the new fermions are given by 
\begin{equation}
L_{Yuk,new}=h H\bar G_R F_L +h_K H_K \bar K_R F_L +c.c. 
\label{6271}
\end{equation}
where $h$ and $h_K$ are the corresponding Yukawa coupling 
parameters, which should be chosen of the order $O(1)$. 
The reader should remember that $v$ (and $v_K$) were assumed 
to be larger than the vevs $v_u$, $v_d$ and $v_e$ which gave masses 
to the ordinary fermions. Therefore with Eq. (\ref{6271}), 
the masses of the new fermions 
are much larger than those of the standard fermions.  

What about mixing terms? Having fixed the set of Higgs multiplets, 
one should in principle write down all possible Yukawa interactions 
which are compatible with the symmetries of the model. In fact 
there exist only two Yukawa interactions in addition to those already 
introduced in Eqs. (\ref{5571}) and (\ref{6271}), namely 
$h_1 H_d \bar G_R q_L$ and $h_2 H_d \bar l_L K_R$, where $h_1$ and 
$h_2$ denote the coupling strenghts. If one adds them to 
Eqs. (\ref{5571}) and (\ref{6271}) and inserts the vacuum expectation 
values, one obtains the complete set of fermion mass terms 
\begin{eqnarray} \nonumber
L_{m}&=&h v\bar U_R U_L +h_K v_K \bar E_R E_L 
     +h_e v_e \bar e_L e_R +h_u v_u \bar u_L u_R
     +h_d v_d \bar d_L d_R
\\  & &
     +h_1 v_d \bar U_R u_L+h_2 v_d \bar e_L E_R
     + c.c.    
\label{9153}
\end{eqnarray}
In this equation we have introduced the notation $F_L=(U_L,E_L)$, 
$G_R=U_R$ and $K_R=E_R$. One concludes from Eq. (\ref{9153}) that 
the mass eigenstates are $d$, $e'$, $E'$, $u'$ and $U'$, 
where $e'$ and $E'$ are linear combinations of $e=e_L+e_R$ 
and $E=E_L+E_R$, and $u'$ and $U'$ are linear combinations of $u=u_L+u_R$ 
and $U=U_L+U_R$. However,  working in the limit that  
$v$ and $v_K$  are larger than $v_u$, $v_d$ and $v_e$, 
the mixing angle is small, 
so that roughly $U' \sim U$, $E' \sim E$, $u' \sim u$ and $e' \sim e$. 
Therefore, the inclusion of the new fermions influences 
the sector of the standard fermions only marginally.   

Another point to discuss is that,  
introducing a new Higgs field $H_K$ with a vev that breaks $SU(4)$, 
one has to make sure that the previously derived properties 
of the symmetry breaking 
in the gauge sector are not spoiled. We have analyzed this 
problem and found that the only modifications induced by $H_K$ 
concern the mass formulae for the leptoquark and for the heavy neutral 
state $T$. All other features of the dynamical symmetry breaking, 
like the relations for the photon and the mass ratio ${m_W \over m_Z}$, 
remain intact. 
The modified leptoquark mass is given by 
\begin{equation}                                  
{1\over 2}m_V^2={1\over 4}
g_4^2(v^2+v_K^2+v_u^2+v_e^2)
\label{6231}
\end{equation}
whereas the axigluon mass is not modified. Thus  
the additional Higgs field $H_K$ tends to increase the leptoquark mass, 
although mass values below 1 TeV are still consistent with all 
requirements. 
Similarly, $H_K$ induces an 
increase in the value of ${1\over 2}m_T^2$ by an amount 
${3\over 8} g_4^2 v_K^2$.  
   
The actual mass 
values of the new fermions can be chosen rather freely by the choice of the 
Yukawa coupling. However, large Yukawa cuplings of order $O(1)$ are more 
appropriate than small ones, because the new fermion masses are 
$hv$ and $h_K v_K$, respectively and should be roughly of the order 
of the leptoquark mass $m_V \approx g_4 
\sqrt{ {1 \over 2} (v^2+v_K^2) }$ 
where $g_4 \sim 1$ and $v \sim v_K$. 

At this point 
it should perhaps be stressed that, even including the 
new fermion multiplets, our model is not as complicated as it may 
appear. It is essentially the Standard Model with 
the main modification that the right--handed up--type 
quarks and leptons form a $SU(4)$ representation. 
All other features of the model then follow from consistency 
requirements, like gauge principle, universality, anomaly 
cancellation etc. 

Cancelling the anomalies by new fermions is a rather ad--hoc 
procedure (although quite common in model building).   
It would be more interesting if the anomalies could be cancelled 
by embedding the $SU(4)\times SU(3)  \times SU(2)_L \times U(1)_X $ 
in an anomaly free grand unified theory. Unfortunately, the group 
is so large, that only GUT groups of rank $\geq 7$ like $E_7$, 
$E_8$ or $SO(18)$ with inconveniently large fermion 
representations are possible. Therefore, unification is not 
a straightforward option in our model. In any case, the new 
particles -- leptoquarks, axigluons etc. -- 
modify the running of coupling constants, so that the GUT 
scenario is strongly modified by the additional $SU(4)$ symmetry. 

{\bf Summary. }
We believe that our model has general implications on gauge
models with vector leptoquarks of mass $\leq$ 1 TeV.
Therefore as a summary we want to stress its general features, 
which are independent of the chosen symmetry breaking mechanism. 
One of them is the appearance of an SU(4) quartet $(u_R,e_R)$ with 
couplings to leptoquarks. 
Any other combination, involving e.g. 
d--quarks, would be in conflict with low energy constraints. 
 
Among the SU(4) gauge bosons there are, besides leptoquarks, 
necessarily neutral as well as gluon--like fields whose masses 
are of the order of the leptoquark mass.  
This is enforced in order to close the color algebra. The 
gluonic type gauge bosons will mix with the ordinary gluons, 
and relations of the
form $g_s=g_{LQ} \cos \theta$ will appear through this mixing
forcing the leptoquark coupling to be a strong coupling,
$g_{LQ} \geq g_s$.

Within the proposed symmetry breaking scheme, it has turned out 
that there is one Higgs field which strongly resembles the 
Standard Model Higgs particle. Further, there is another one 
which is mainly responsible for the breaking of the SU(4) 
symmetry.  

In the present model, the right--handed up--type quarks 
are part of the SU(4) quartets whereas the right--handed down--type 
quarks are SU(4) singlets. 
It is thus apparent that the custodial symmetry between 
$u_R$ and $d_R$ which is respected by the Standard Model 
(neglecting quark masses) is violated.   
Correspondingly, one expects "large" loop corrections to 
the $\rho$--parameter \cite{veltman} (or $\epsilon_{1,2,3}$ 
\cite{altarelli1}) other than $\sim m_t^2-m_b^2$. In fact, 
there are additional self--energy diagrams of W and Z 
with either leptoquarks $V^{\pm}$,  
the neutral Z' ($T_{\mu}$) as well as many of the Higgs components 
discussed above. In addition, there are the new fermions needed 
for anomaly cancellation. Of course one can always argue 
that the contribution of these particles to $\epsilon_{1,2,3}$ 
is sufficiently small 
if their masses are larger than, say, 500 GeV 
\cite{altarelli1}. 

{\bf Appendix. }
Finally we want to discuss what happens if the assumption of 
vanishing fermion number is given up. Low energy constraints, 
universality and gauge principle then allow for a leptoquark 
interaction of the form $\bar u_R^c \gamma^{\mu} l_L U_{\mu}$,  
where $U_{\mu}$ is an $SU(2)_L$ doublet of vector leptoquarks. 
Using the same philosophy as before, one may now construct 
a gauge theory based on 
$SU(5) \times SU(3)  \times SU(2)_q \times U(1)_X $ where 
the left--handed quarks $q_L$ transform as a doublet under $SU(2)_q$ 
whereas the left--handed leptons are part of a $SU(5)$ quintet \footnote
{Note that the quintet is reminiscent of the quintet in "flipped $SU(5)$" 
\cite{antoniadis}. 
However, in contrast to flipped $SU(5)$, here one has no 10--representation 
containing d--quarks because this would induce proton decay.},   
$p_L \equiv (u_R^{1 c},u_R^{2 c},u_R^{3 c},\nu_L,e_L)$.   
Thus the $SU(5)$ 
contains a subgroup $SU(3)' \times SU(2)_l$ where 
$SU(3)' \times SU(3)\rightarrow SU(3)_c$ as before, 
giving rise to 8 massive axigluons, and 
$SU(2)_l\times SU(2)_q\rightarrow SU(2)_L$, introducing 3 
additional massive states, the "axi--$W^{\pm}/Z$". 
There is also a neutral gauge boson S which mixes with 
the photon. 
The remaining 12 $SU(5)$ gauge bosons constitute the leptoquark $U_{\mu}$. 

\begin{table} 
\label{tab4}  
\begin{center}
\begin{tabular}{|c|c|c|c|c|}
\hline
 & $SU(5)$ & $SU(3)$ & $SU(2)_q$ & $U(1)_X$ \\
\hline   
$q_L$ & 1 & 3 & 2 & ${1\over 6}$ \\
$p_L$ & 5 & 1 & 1 & ${1\over 5}$ \\
$e_R$ & 1 & 1 & 1 & $-1$ \\
$d_R$ & 1 & 3 & 1 & $-{1\over 3}$ \\   
\hline
\end{tabular}   
\bigskip
\caption{Quantum number assignments of the $F=-2$ leptoquark model}
\end{center}
\end{table}

To give some more details of the model, the quantum number 
assignments of the standard fermions are shown in Table 4. 
The $U(1)$ charge of the quintet can be fixed to be 1/5 by the requirement that 
the photon coupling is vectorlike. This as well as many other 
features of the model work out in the same way as the $SU(4)$ 
model presented in the main text. For example, one now has 
\begin{equation}
Q=X-{7\over \sqrt{15}} T_{24} +T_3^q +T_3^l
\label{50911}
\end{equation}
where $T_{24}={1\over \sqrt{15}}$diag$(-1,-1,-1,{3\over 2},{3\over 2})$ 
and $T_3^l$ and $T_3^q$ are the diagonal generators of $SU(5)$ and $SU(2)_q$, 
respectively. Note that the $SU(5)$ gauge bosons $R_{\mu}$ are decomposed 
into $SU(3)'$ gauge bosons $R_{\mu}'$, $SU(2)_l$ gauge bosons 
$W_{\mu}^l$, the leptoquarks $U_{\mu}$ and a singlet $S_{\mu}$, which 
is related to the $T_{24}$ generator. More precisely, one has 
\begin{equation}
R_{\mu}= {1\over \sqrt{2}}
\pmatrix{ R_{\mu}' -{ 2\over \sqrt{30}} S_{\mu}\times 1_3  &U_{\mu}\cr
        U^+_{\mu} & W_{\mu}^l +{ 3\over \sqrt{30}}S_{\mu}\times 1_2 \cr} \, .
\label{55732}        
\end{equation}
The group $SU(5) \times SU(3)  \times SU(2)_q$ can be broken to 
$SU(3)_c \times SU(2)_L$ by two Higgs multiplets, 
$H_3(\bar 5,3,1,-{7\over 15})$ with vev $v_3 \delta_{\bar \alpha \beta}$ 
and $H_2(\bar 5,1,2,{7\over 10})$ with vev $v_2 \delta_{\bar \alpha i}$, 
where $\alpha$, $\beta$ and $i$ denote $SU(5)$, $SU(3)$ and $SU(2)_q$  
indices, respectively. The following vector boson masses are then 
obtained: 
\begin{itemize}
\item Axigluon mass : ${1\over 2}m_N^2={1\over 2}
(g_3^2+g_5^2) v_3^2$ 
\item Axi--$W^{\pm}/Z$ mass : ${1\over 2}m_M^2={1\over 2}
(g_{2q}^2+g_5^2) v_2^2$ 
\item Leptoquark mass : ${1\over 2}m_U^2={1\over 4}
g_5^2(v_2^2+v_3^2)$  
\item Neutral vector boson mass : ${1\over 2}m_S^2={1\over 2} 
g_5^2({3\over 5}v_2^2+{2\over 5}v_3^2)$  
\end{itemize} 
In these expressions, $g_5$, $g_3$ and $g_{2q}$ denote the 
couplings of $SU(5)$, $SU(3)$ and $SU(2)_q$, respectively. 
Note that one has 
\begin{equation}
g_s^{-2}=g_3^{-2}+g_5^{-2}  
\label{557431}
\end{equation}
\begin{equation}
g_2^{-2}=g_{2q}^{-2}+g_5^{-2}                           
\label{557432}
\end{equation}
and
\begin{equation}
g_Y^{-2}=g_X^{-2}+{49 \over 15}g_5^{-2}
\label{557932}
\end{equation}
where $g_s$, $g_2$ and $g_Y$ are the couplings of the Standard Model gauge group 
$SU(3)_c \times SU(2)_L \times U(1)_Y$. 
Eqs. (\ref{557431}), (\ref{557432}) and (\ref{557932}) 
are in analogy to Eqs. (\ref{5574}) and (\ref{491711}) 
for the $SU(4)$--type model. 
They imply that in general $g_{2q} << g_{3,5}$. This suggests 
that the axi--$W^{\pm}/Z$ mass might be somewhat smaller than the 
axigluon mass, although this is not compelling because 
the ratio of these masses depends also on the relative magnitude 
of $v_2$ and $v_3$. 

As in the $SU(4)$ model, the 
fermion masses $m_e$, $m_u$ and $m_d$ 
can be obtained from three Higgs fields,  
$H_e(5,1,1,{6\over 5})$, $H_u(5,3,2,{11\over 30})$ and 
$H_d(1,1,2,{1\over 2})$ with vevs 
$v_e \delta_{\alpha 5}$, $v_u \delta_{\alpha \beta} \delta_{i 1}$ 
and $v_d \delta_{i 2}$.   
These expectation values in principle contribute to the 
vector boson masses as well, so that a complicated mixing 
matrix as in Eq. (\ref{5577}) arises. However, just as 
in the $SU(4)$ model, it turns out that $v_{e,u,d} << v_{2,3}$, 
so that the mass formulas given before Eq. (\ref{557431})  
are still approximately valid (up to terms of order $v_{e,u,d}^2$).   
As in the $SU(4)$ model, the
vevs $v_{e,u,d}$ determine not only the fermion masses but also 
the mass of the W-- and Z--boson, where by use of Eq. (\ref{50911}) 
one is lead to the correct value of the Weinberg angle.

{\bf Acknowledgement. }                          
We are indebted to Bill Bardeen for some helpful suggestions and 
to Sacha Davidson for many informations regarding her paper.

\end{document}